\documentclass[prd,twocolumn,superscriptaddress,showpacs,amssymb,amsmath,
amsfonts,aps,altaffilletter]{revtex4}

\usepackage{color}
\usepackage{times}
\usepackage{graphicx}
\usepackage{fancyhdr}
\usepackage{float}
\usepackage{ulem}
\makeatletter
\makeatletter
\def\@fnsymbol#1{\ifcase#1\or * \or  $+$ \or  \$ \or \#  \or \dag \or \ddag \or
$\mathsection$ \or $ \mathparagraph$ \or $\|$  \or \textordfeminine \or \textbul
let   
\or ** \or $++$ \or  \$\$ \or \#\#  \or \dag\dag \or \ddag\ddag \or
$\mathsection\mathsection$ \or $ \mathparagraph\mathparagraph$ \or $\|\|$  \or 
\textordfeminine\textordfeminine \or \textbullet \textbullet \or *** \or $+++$ 
\or  \$\$\$ \or \#\#  \or \dag\dag \or \ddag\ddag \or
$\mathsection \mathsection\mathsection$ \or $ \mathparagraph 
\mathparagraph\mathparagraph$ \or $\|\|\|$  \or 
\textordfeminine\textordfeminine\textordfeminine \or 
\textbullet\textbullet\textbullet \or \else \@ctrerr\fi}
\makeatother

\newcommand\unit[2]{\mbox{#1\,#2}}
\newcommand\mathunit[2]{#1\,#2}
\def\thercsid{\relax}
\def\rcsid#1{\def\next##1#1{\def\thercsid{##1}}\next}
\rcsid$Id: ringdown.tex,v 1.54 2009/08/17 13:30:56 lgoggin Exp $

\renewcommand{\today}{\number\day\space\ifcase\month\or
  January\or February\or March\or April\or May\or June\or
  July\or August\or September\or October\or November\or December\fi
  \space\number\year}

\begin{document}

\title{Search for gravitational wave ringdowns from perturbed black holes in LIGO S4 data}

\newcommand*{\AG}{Albert-Einstein-Institut, Max-Planck-Institut f\"{u}r Gravitationsphysik, D-14476 Golm, Germany}
\affiliation{\AG}
\newcommand*{\AH}{Albert-Einstein-Institut, Max-Planck-Institut f\"{u}r Gravitationsphysik, D-30167 Hannover, Germany}
\affiliation{\AH}
\newcommand*{\AU}{Andrews University, Berrien Springs, MI 49104 USA}
\affiliation{\AU}
\newcommand*{\AN}{Australian National University, Canberra, 0200, Australia}
\affiliation{\AN}
\newcommand*{\CH}{California Institute of Technology, Pasadena, CA  91125, USA}
\affiliation{\CH}
\newcommand*{\CA}{Caltech-CaRT, Pasadena, CA  91125, USA}
\affiliation{\CA}
\newcommand*{\CU}{Cardiff University, Cardiff, CF24 3AA, United Kingdom}
\affiliation{\CU}
\newcommand*{\CL}{Carleton College, Northfield, MN  55057, USA}
\affiliation{\CL}
\newcommand*{\CS}{Charles Sturt University, Wagga Wagga, NSW 2678, Australia}
\affiliation{\CS}
\newcommand*{\CO}{Columbia University, New York, NY  10027, USA}
\affiliation{\CO}
\newcommand*{\ER}{Embry-Riddle Aeronautical University, Prescott, AZ   86301 USA}
\affiliation{\ER}
\newcommand*{\EU}{E\"{o}tv\"{o}s University, ELTE 1053 Budapest, Hungary}
\affiliation{\EU}
\newcommand*{\HC}{Hobart and William Smith Colleges, Geneva, NY  14456, USA}
\affiliation{\HC}
\newcommand*{\IA}{Institute of Applied Physics, Nizhny Novgorod, 603950, Russia}
\affiliation{\IA}
\newcommand*{\IU}{Inter-University Centre for Astronomy  and Astrophysics, Pune - 411007, India}
\affiliation{\IU}
\newcommand*{\HU}{Leibniz Universit\"{a}t Hannover, D-30167 Hannover, Germany}
\affiliation{\HU}
\newcommand*{\CT}{LIGO - California Institute of Technology, Pasadena, CA  91125, USA}
\affiliation{\CT}
\newcommand*{\LO}{LIGO - Hanford Observatory, Richland, WA  99352, USA}
\affiliation{\LO}
\newcommand*{\LV}{LIGO - Livingston Observatory, Livingston, LA  70754, USA}
\affiliation{\LV}
\newcommand*{\LM}{LIGO - Massachusetts Institute of Technology, Cambridge, MA 02139, USA}
\affiliation{\LM}
\newcommand*{\LU}{Louisiana State University, Baton Rouge, LA  70803, USA}
\affiliation{\LU}
\newcommand*{\LE}{Louisiana Tech University, Ruston, LA  71272, USA}
\affiliation{\LE}
\newcommand*{\LL}{Loyola University, New Orleans, LA 70118, USA}
\affiliation{\LL}
\newcommand*{\MT}{Montana State University, Bozeman, MT 59717, USA}
\affiliation{\MT}
\newcommand*{\MS}{Moscow State University, Moscow, 119992, Russia}
\affiliation{\MS}
\newcommand*{\ND}{NASA/Goddard Space Flight Center, Greenbelt, MD  20771, USA}
\affiliation{\ND}
\newcommand*{\NA}{National Astronomical Observatory of Japan, Tokyo  181-8588, Japan}
\affiliation{\NA}
\newcommand*{\NO}{Northwestern University, Evanston, IL  60208, USA}
\affiliation{\NO}
\newcommand*{\RI}{Rochester Institute of Technology, Rochester, NY  14623, USA}
\affiliation{\RI}
\newcommand*{\RA}{Rutherford Appleton Laboratory, HSIC, Chilton, Didcot, Oxon OX11 0QX United Kingdom}
\affiliation{\RA}
\newcommand*{\SJ}{San Jose State University, San Jose, CA 95192, USA}
\affiliation{\SJ}
\newcommand*{\SM}{Sonoma State University, Rohnert Park, CA 94928, USA}
\affiliation{\SM}
\newcommand*{\SE}{Southeastern Louisiana University, Hammond, LA  70402, USA}
\affiliation{\SE}
\newcommand*{\SO}{Southern University and A\&M College, Baton Rouge, LA  70813, USA}
\affiliation{\SO}
\newcommand*{\SA}{Stanford University, Stanford, CA  94305, USA}
\affiliation{\SA}
\newcommand*{\SR}{Syracuse University, Syracuse, NY  13244, USA}
\affiliation{\SR}
\newcommand*{\PU}{The Pennsylvania State University, University Park, PA  16802, USA}
\affiliation{\PU}
\newcommand*{\UM}{The University of Melbourne, Parkville VIC 3010, Australia}
\affiliation{\UM}
\newcommand*{\MI}{The University of Mississippi, University, MS 38677, USA}
\affiliation{\MI}
\newcommand*{\SF}{The University of Sheffield, Sheffield S10 2TN, United Kingdom}
\affiliation{\SF}
\newcommand*{\TA}{The University of Texas at Austin, Austin, TX 78712, USA}
\affiliation{\TA}
\newcommand*{\TC}{The University of Texas at Brownsville and Texas Southmost College, Brownsville, TX  78520, USA}
\affiliation{\TC}
\newcommand*{\TR}{Trinity University, San Antonio, TX  78212, USA}
\affiliation{\TR}
\newcommand*{\BB}{Universitat de les Illes Balears, E-07122 Palma de Mallorca, Spain}
\affiliation{\BB}
\newcommand*{\UA}{University of Adelaide, Adelaide, SA 5005, Australia}
\affiliation{\UA}
\newcommand*{\BR}{University of Birmingham, Birmingham, B15 2TT, United Kingdom}
\affiliation{\BR}
\newcommand*{\FA}{University of Florida, Gainesville, FL  32611, USA}
\affiliation{\FA}
\newcommand*{\GU}{University of Glasgow, Glasgow, G12 8QQ, United Kingdom}
\affiliation{\GU}
\newcommand*{\MD}{University of Maryland, College Park, MD 20742 USA}
\affiliation{\MD}
\newcommand*{\AM}{University of Massachusetts - Amherst, Amherst, MA 01003, USA}
\affiliation{\AM}
\newcommand*{\MU}{University of Michigan, Ann Arbor, MI  48109, USA}
\affiliation{\MU}
\newcommand*{\MN}{University of Minnesota, Minneapolis, MN 55455, USA}
\affiliation{\MN}
\newcommand*{\OU}{University of Oregon, Eugene, OR  97403, USA}
\affiliation{\OU}
\newcommand*{\RO}{University of Rochester, Rochester, NY  14627, USA}
\affiliation{\RO}
\newcommand*{\SL}{University of Salerno, 84084 Fisciano (Salerno), Italy}
\affiliation{\SL}
\newcommand*{\SN}{University of Sannio at Benevento, I-82100 Benevento, Italy}
\affiliation{\SN}
\newcommand*{\SH}{University of Southampton, Southampton, SO17 1BJ, United Kingdom}
\affiliation{\SH}
\newcommand*{\SC}{University of Strathclyde, Glasgow, G1 1XQ, United Kingdom}
\affiliation{\SC}
\newcommand*{\WA}{University of Western Australia, Crawley, WA 6009, Australia}
\affiliation{\WA}
\newcommand*{\UW}{University of Wisconsin-Milwaukee, Milwaukee, WI  53201, USA}
\affiliation{\UW}
\newcommand*{\WU}{Washington State University, Pullman, WA 99164, USA}
\affiliation{\WU}

\author{}    \affiliation{\GU}    
\author{B.~P.~Abbott}    \affiliation{\CT}    
\author{R.~Abbott}    \affiliation{\CT}    
\author{R.~Adhikari}    \affiliation{\CT}    
\author{P.~Ajith}    \affiliation{\AH}    
\author{B.~Allen}    \affiliation{\AH}  \affiliation{\UW}  
\author{G.~Allen}    \affiliation{\SA}    
\author{R.~S.~Amin}    \affiliation{\LU}    
\author{S.~B.~Anderson}    \affiliation{\CT}    
\author{W.~G.~Anderson}    \affiliation{\UW}    
% ...
 \collaboration{The LIGO Scientific Collaboration, http://www.ligo.org}
 \noaffiliation
%
%
%
%
%
%

\date[\relax]{ RCS \thercsid; compiled \today }

\begin{abstract}\quad
According to general relativity a perturbed black hole will settle to a
stationary configuration by the emission of gravitational radiation. Such a perturbation will occur, for example, in the coalescence of a black hole binary, following their inspiral and subsequent merger. At late times the waveform is a superposition of quasi-normal modes, which we refer to as the ringdown. The dominant mode is expected to be the fundamental mode, $l=m=2$. Since this is a well-known waveform, matched filtering can be implemented to search for this signal using LIGO data. We present a search for gravitational waves from black hole ringdowns in the fourth LIGO science run S4, during which LIGO was sensitive to the dominant mode of perturbed black holes with masses in the range of \unit{10}{$M_\odot$} to \unit{500}{$M_\odot$}, the regime of intermediate-mass black holes, to distances up to \unit{300}{Mpc}. We present a search for gravitational waves from black hole ringdowns using data from S4. No gravitational wave candidates were found; we place a 90\%-confidence upper limit on the rate of ringdowns from black holes with mass between \unit{85}{$M_\odot$} and \unit{390}{$M_\odot$} in the local universe, assuming a uniform distribution of sources, of \unit{$3.2\times 10^{-5}$}{yr$^{-1}$ Mpc$^{-3}$} $=$ \unit{$1.6 \times 10^{-3}$}{yr$^{-1}$ $L_{10}^{-1}$}, where $L_{10}$ is $10^{10}$ times the solar blue-light luminosity.
\end{abstract}
\maketitle

\section{Introduction}
\label{sec:intro}

The existence of intermediate mass black holes, IMBHs, ($20 M_\odot \leq M \leq 10^6 M_\odot$) has been under debate for several decades. While general relativity does not preclude IMBHs, there had been no observational evidence for their existence until recently.
Electromagnetic observations have indicated that ultra-luminous X-ray sources, that is, sources radiating above the Eddington luminosity for a stellar mass black hole, may be powered by IMBHs. Strong evidence in support of this argument has recently been reported with the discovery of a source whose luminosity implies the presence of black hole with mass of at least \unit{500}{$M_\odot$}~\cite{2009Natur.460...73F}. Further information pertaining to IMBHs may be found in recent comprehensive review articles~\cite{2004IJMPD..13....1M,miller-2008}.

Predictions have been made for the rate of detection of ringdowns from IMBHs in Advanced LIGO. Ref.~\cite{imbhlisa-2006} predicts a rate of 10 events per year from IMBH-IMBH binary coalescences. When scaled to the sensitivity of the data set under consideration in this investigation, this prediction becomes \unit{$10^{-4}$}{yr$^{-1}$} \cite{Goggin:2008}. Ringdowns following coalescences of stellar-mass BHs with IMBHs could also be detectable with Advanced LIGO, with possible rates of tens of events per year \cite{Mandel:2007rates}.

Detection of gravitational radiation from IMBHs however, would provide unambiguous evidence of their existence. In order for such an object to reveal itself through gravitational radiation it must come to be in a perturbed state, for example as the remnant of the coalescence of two IMBHs.  Current ground-based gravitational wave detectors, such as the Laser Interferometer Gravitational-Wave Observatory (LIGO), operate in an optimal frequency range for the detection of the ringdown phase of the binary coalescence of IMBH binaries. In this paper we describe a search for ringdown waveforms in data from the fourth LIGO science run, S4.

\section{The Ringdown Waveform}
A series of studies within a linearized approximation to Einstein's equations and also full-blown numerical simulations have shown that the gravitational wave signal from a perturbed black hole consists of roughly three stages~\cite{2009CQGra..26p3001B}: (i) A prompt response at early times, which depends strongly on the initial conditions, and is the counterpart to light-cone propagation; (ii) An exponentially decaying ``ringdown'' phase at intermediate times, where  quasi-normal modes, QNMs, dominate the signal, which depends entirely on the final black hole's parameters; (iii) A late-time tail, usually a power-law fall-off of the field. The ringdown phase, which is the focus of this work, starts roughly when the perturbing source reaches the peak of the potential barrier around the black hole, and consists of a super-position of quasinormal modes. For instance, during the merger of two black holes, the start of the ringdown is roughly associated with the formation of a common apparent horizon, which also corresponds to the peak of the gravitational-wave amplitude. For black holes in the LIGO band this is on the order of tens of milliseconds after the innermost stable circular orbit of the binary.
Each quasi-normal mode has a characteristic complex angular frequency $\omega_{lm}$; the real part is the angular frequency and the imaginary part is the inverse of the damping time $\tau$. Numerical simulations (for example~\cite{buonanno:124018}) have demonstrated that the dominant mode is the fundamental mode, $l=m=2$, and that far from the source the waveform can be approximated by
\begin{equation}
h_0(t)= \Re \left\{ \mathcal{A}\frac{GM}{c^2r} \mathrm{e}^{-\mathrm{i} \omega_{22} t}\right\},
\label{eqn:waveform}
\end{equation}
where $\mathcal{A}$ is the dimensionless amplitude of the $l=m=2$ mode, $r$ is the distance to the source, $M$ is the black hole mass, $c$ is the speed of light,  and $G$ is the gravitational constant. This is usually expressed in terms of the oscillation frequency $f_0=\Re\left(\omega_{22}\right)/2\pi$ and the quality factor $Q=\pi f_0 / \Im\left(\omega_{22}\right)$, 
\begin{equation}
h_0(t)=\mathcal{A}\frac{GM}{c^2r} \mathrm{e}^{-\pi f_0 t / Q } \cos \left( 2 \pi f_0 t\right).
\end{equation}
Under the assumption that the waveform is completely known, we can implement the method of matched filtering~\cite{wainstein:1962}, in which the data is correlated with a bank of signal templates parametrized by the ringdown frequency and quality factor.
An analytic fit by Echeverria~\cite{Echeverria:1989hg} to Leaver's numerical calculations~\cite{Leaver:1985ax} relates the waveform parameters to the black hole's physical parameters, mass $M$, and dimensionless spin factor, defined in terms of the spin angular momentum $J$, for the fundamental mode $\hat{a}=Jc/GM^2$:
\begin{eqnarray}
f_{0}&=&\frac{1}{2 \pi} \frac{c^3}{GM} g(\hat{a})  \label{eqn:Echeverria_fofMa} \\
Q&=&2\left( 1-\hat{a}\right) ^{-9/20},
\label{eqn:Echeverria_Qofa}
\end{eqnarray}
where $g(\hat{a}) = 1-0.63 \left( 1-\hat{a} \right) ^{3/10}$. Thus, if we detect the $l=m=2$ mode, these formulae will provide the mass and spin of the black hole~\footnote{Recent work by Berti et al.~\cite{2006PhRvD..73f4030B} provides analytic expressions for higher order modes.}.

The amplitude is given by~\cite{Goggin:2008}
\begin{equation}
\mathcal{A}=\sqrt{\frac{5}{2}\epsilon} \:
Q^{-\frac{1}{2}}F(Q)^{-\frac{1}{2}} g(a)^{-\frac{1}{2}}
\label{eqn:amplitude}
\end{equation}
where $F(Q)= 1+\frac{7}{24 Q^2}$.
In addition to a frequency and quality factor dependence, the amplitude of the waveform also depends on the fraction of the final black hole's mass radiated as gravitational waves, $\epsilon$. This quantity scales with the square of the symmetric mass ratio $\eta$, where $\eta=m_1m_2/(m_1+m_2)^2$, and thus is largest for an equal mass binary \cite{flanagan-1998-57,Berti:2007}. Numerical simulations of the merger of equal mass binaries have shown that approximately 1\% of the final black hole's mass is emitted in gravitational waves~\cite{buonanno:124018}. In this search we do not attempt to evaluate $\epsilon$; we use the output of the filter to calculate the effective distance to a source emitting 1\% of its mass as gravitational waves. The effective distance is the distance to an optimally located and oriented source.


\section{Data Set}
\begin{figure}[htb]
\centering
\begin{center}
\includegraphics[width=0.3\textwidth]{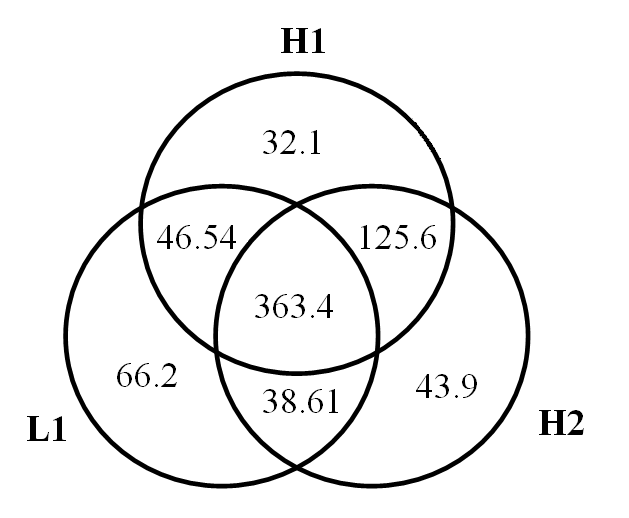}
\caption{Venn diagram of the coincident detector times in hours.}
\label{fig:venn}
\end{center}
\end{figure}

This search uses data from the 4$^{\textrm{th}}$ LIGO science run (S4), which took place between February 22$^{\textrm{nd}}$ and March 24$^{\textrm{th}}$, 2005. This yielded a total of \unit{567.4}{hours} of analyzable data from the \unit{4}{km} interferometer in Hanford, WA (H1), \unit{571.3}{hours} from the \unit{2}{km} interferometer in Hanford, WA (H2), and \unit{514.7}{hours} from the \unit{4}{km} interferometer in Livingston, LA (L1). In this analysis we require that data be available from at least two detectors at any given time. This results in approximately \unit{364}{hours} of triple coincidence and \unit{210}{hours} of double coincidence, as shown in Fig.~\ref{fig:venn}. During S4, the LIGO detectors operated significantly below their design sensitivity; this was attained in the subsequent science run, S5~\cite{abbott-2007}.

\begin{figure}[htb]
\centering
\begin{center}
\includegraphics[width=0.5\textwidth]{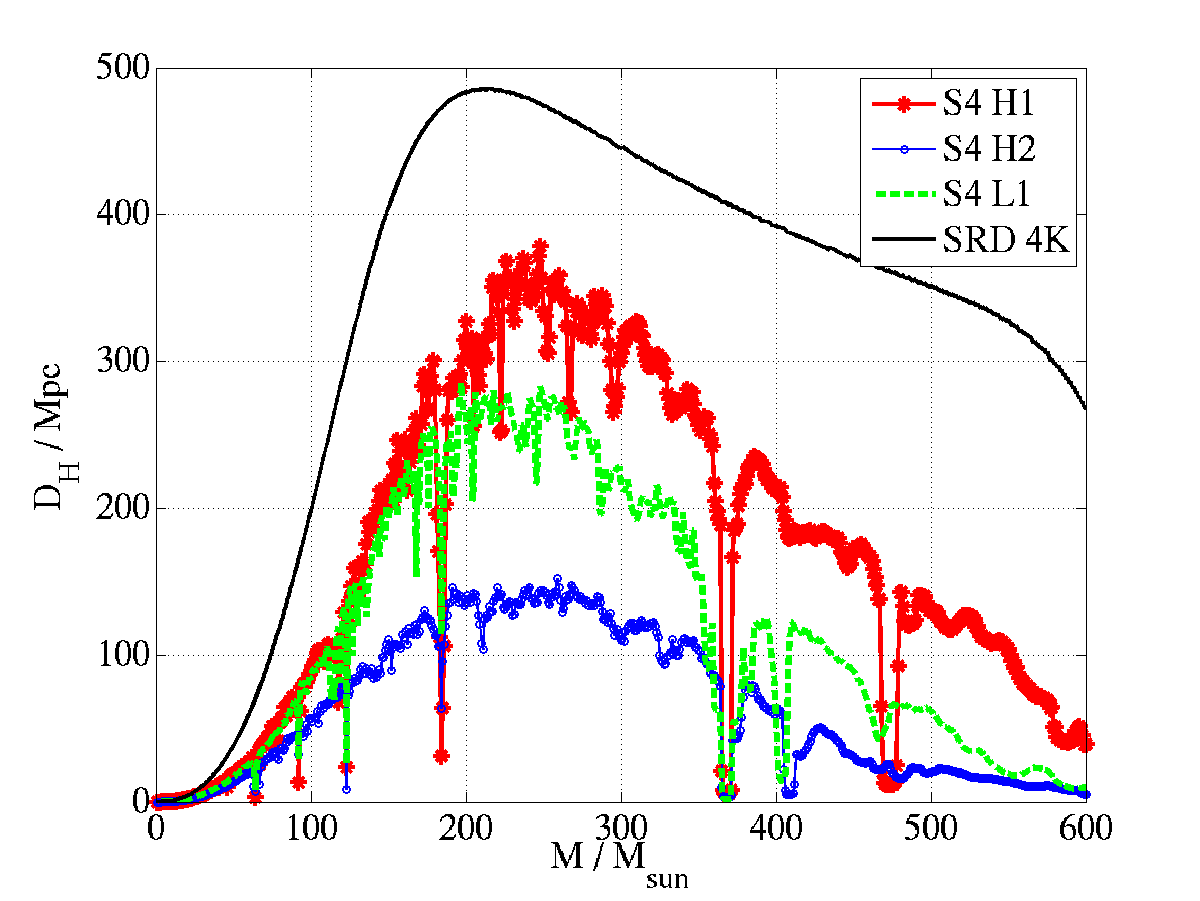}
\caption{The horizon distance versus mass and frequency for a black hole with spin of $\hat{a}=0.9$ and $\epsilon=0.01$. From top to bottom, the curves show the horizon distance for the Initial LIGO reference design (black), the Hanford 4 km detector H1 (red), the Livingston 4 km detector L1 (green), and the Hanford 2 km detector H2 (blue).}
\label{fig:horiz}
\end{center}
\end{figure}

The LIGO detectors are sensitive to gravitational waves in the frequency band of $\sim$\unit{50}{Hz} to \unit{2}{kHz}. This corresponds to a mass range of 11 to \unit{440}{$M_\odot$} for a black hole with $\hat{a}=0.9$ oscillating in its fundamental mode. Using a typical instantiation of the S4 noise power spectrum we can estimate the horizon distance, $D_{\textrm{H}}$, the distance out to which a specified source with optimal location and orientation produces an SNR of 8 in the detector. We consider a spinning black hole with $\hat{a}=0.9$ and $\epsilon=1\%$. This quantity is shown as a function of mass, for each of the three LIGO detectors and the LIGO design sensitivity in Fig.~\ref{fig:horiz}.

During each science run there are times when disturbances couple into the data stream, introducing noise transients that can trigger a matched filter with high SNR. A careful study of correlations between LIGO data and auxiliary channels allowed us to identify data segments with an excessive noise transient rate, or with known artifacts. We refer to these as data quality flags. These are then categorized according to the severity of the disturbance~\cite{LIGOS3S4Tuning}; triggers occurring during category 1 times are not analyzed as the excess of noise is likely to contaminate the estimation of the power spectral density. Data occurring during category 2 or 3 times are less problematic, and thus to avoid excessive segmentation of the data, these are vetoed during the analysis. Any gravitational wave candidates occurring during category 4 times are cautiously examined. 

\section{Pipeline} \label{sec:pipeline}
When the signal is known, the optimal method of extracting the signal from Gaussian noise is matched filtering. LIGO data is non-Gaussian and while the method is still appropriate it is not sufficient to discriminate between signal and background. We employ an analysis pipeline which was designed for this purpose, and closely resembles that of the inspiral searches described in~\cite{brown-2005-22,LIGOS3S4all}. Here we summarize the main steps. The first stage of the pipeline involves reading in and conditioning the data $s(t)$ from each of the three LIGO detectors. We read in uncalibrated data, the differential arm length signal, at a sample rate of \unit{16384}{s$^{-1}$} and down-sample to \unit{8192}{s$^{-1}$}. This is converted to strain by applying the detector response function. The one-sided power spectral density $S_h(f)$ is calculated for each \unit{2176}{s} long segment of the calibrated data. The data is then broken further into sets of sixteen overlapping blocks, \unit{256}{s} in length, and filtered using a bank of ringdown templates.
The templates are positioned in $f_0$ and $Q$ according to the metric~\cite{Creighton:1999pm}
\begin{eqnarray} \mathrm{d}s^2 &=& \frac{1}{8} \Bigg[ \frac{3+16Q^4}{Q^2(1+4Q^2)^2} \:
                        \mathrm{d} Q^2 - 2 \frac{3+4Q^2}{f_0 Q(1+4Q^2)} \:
                        \mathrm{d} Q \:\mathrm{d} f_0 \nonumber \\
                      & & \qquad \qquad +\frac{3+8Q^2}{f_0^2} \: \mathrm{d} f_0^2 \Bigg],
\label{eqn:metric}
\end{eqnarray}
such that the maximum mismatch, d$s^2$, between any point within the template bank and the nearest template is 3\%. We search within the most sensitive portion of the LIGO frequency band, \unit{50}{Hz} to \unit{2}{kHz} and in quality factor between 2 and 20. With these parameters we obtain a bank of 583 templates with five different values of $Q$. The same template bank is used throughout the run. Numerical simulations~\cite{rezzolla-2008-679,hinder-2008-77,marronetti-2008-77,dain-2008} have demonstrated that the maximum spin attained by the final black hole in a binary black hole merger is less than 0.96, corresponding to a quality factor of 8.5. In this search, we chose to cover a larger parameter space, and explore spin values between 0 ($Q=2$) and 0.994 ($Q=20$). 
Each filter has the form
\begin{equation}
h(t)=\cos\left(2\pi f_0 t\right) \;\mathrm{e}^{-\pi f_0 t / Q}, \qquad 0\leq t \leq t_\textrm{max}
\end{equation}
with a length of ten e-folding times, $t_{\textrm{max}}=10 \:\tau$, where $\tau=Q/\pi f_0$. Filtering the data gives a signal to noise ratio (SNR)
\begin{equation}
\rho(h)=\frac{\left<s,h\right>}{\sqrt{\left<h,h\right>}},
\end{equation}
where
\begin{equation}
\left<s,h\right>=2\int^\infty_{-\infty} \frac{\tilde{s}(f)\tilde{h}^*(f) }{ S_h(\left|f\right|)} \textrm{d}f,
\end{equation}
Here, the noise spectral density $S_h(f)$ is the one appropriate for the data segment in question.
For each filter, only triggers exceeding a pre-defined threshold, $\rho_{*}=5.5$ are retained. These are then clustered using a peak finding algorithm over a minimum of \unit{1}{s}, with the loudest triggers for each filter written out to a file. Approximately $10^6$ triggers were written out for each detector for the entire S4 run. 

\begin{figure}[htb]
\centering
\begin{center}
\includegraphics[width=0.5\textwidth]{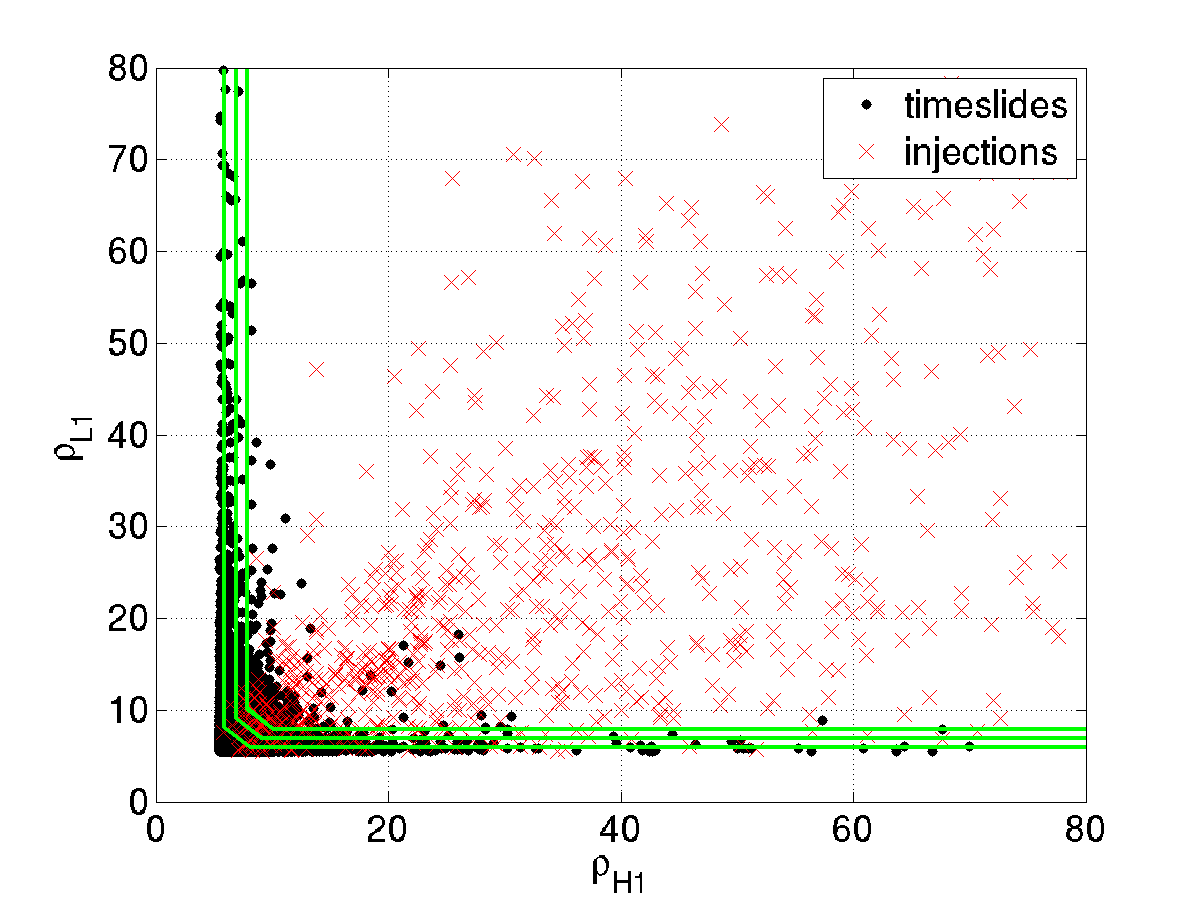}
\caption{Plot of background events (black dot) and simulated signal (red x's) found in double coincidence. The green lines denote the lines of constant detection statistic for two detector coincidences.}
\label{fig:detstat}
\end{center}
\end{figure}

In order to claim a detection of a gravitational wave ringdown we require coincidence between at least two detectors in the time of arrival of the signal and in the waveform parameters. The time requirement is enforced first, only allowing triggers that appear within \unit{4}{ms} of each other for co-located Hanford detectors and \unit{14}{ms} for the Livingston-Hanford pairs (the two observatories are separated by \unit{10}{ms} of light travel time) through to the next stage. The second coincidence test is based on the metric used to lay out the template bank, equation (\ref{eqn:metric}), with the size of the coincidence window around each template varying depending on its position within the parameter space. 
At this stage we also veto triggers occurring during times when category 2 and 3 data quality flags were on, and implement an amplitude consistency test between triggers appearing in both Hanford detectors, such that the pair of triggers is retained only if the ratio of the H1 effective distance to the H2 effective distance is less than 2. This results in list of coincident triggers found in two or three detectors, hereafter referred to as doubles and triples respectively. The final step is to cluster this coincidences over a time window of 10 s, retaining only the coincidences with the highest value of a detection statistic $\rho_{\textrm{DS}}$, a measure of the relative significance of the coincidences.
For triple coincidences (H1H2L1) the detection statistic was
\begin{equation}
\rho_{\textrm{DS}}=\rho_{\textrm{H1}}^2+\rho_{\textrm{H2}}^2+\rho_{\textrm{L1}}^2. \label{eqn:tripDS}
\end{equation}
The high level of false alarms associated with two detector coincidences (H1L1, H1H2, or H2L1), shown in Fig.~\ref{fig:detstat}, warranted a different detection statistic,
\begin{equation}
\rho_{\textrm{DS}}=\textrm{min}\{\rho_{\textrm{ifo1}}+\rho_{\textrm{ifo2}}, \: a \times \rho_{\textrm{ifo1}}+b, \: a \times \rho_{\textrm{ifo2}}+b\},
\end{equation}
where suitable values of $a$ and $b$ were found to be 2 and 2.2 respectively from an evaluation of the false alarm rate. (The evaluation of false alarm rates is discussed in section \ref{sec:tuning}.) Note that in this ranking we do not take the square root of equation (\ref{eqn:tripDS}) so as to emphasize the triple coincidences over the doubles. 
The ten coincidences with the highest value of the detection statistic in each clustered coincidence category (triples or doubles) are followed up upon as detection candidates.

\section{Tuning the Search} \label{sec:tuning}
As with previous matched filtering searches in LIGO we implement a ``blind analysis'' of the data, that is, the constraints are decided upon prior to looking at the full data set. We chose values of the constraints that maximize the number of ringdown signals recovered while minimizing the false alarm rate. The constraints under consideration include the SNR threshold, the size of the coincidence windows, the detection statistic, and the amplitude consistency test.  

\begin{figure}[htb]
\centering
\begin{center}
\includegraphics[width=0.5\textwidth]{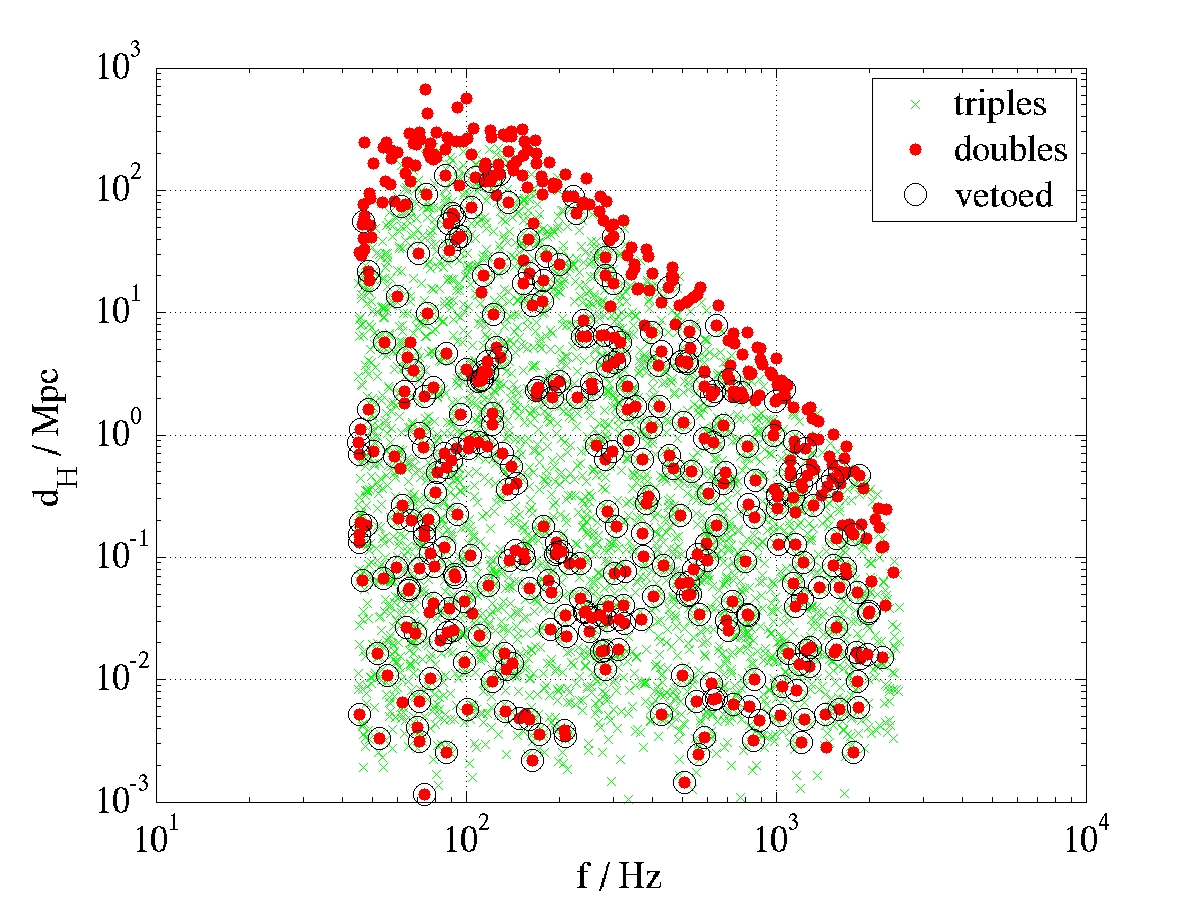}
\caption{Hanford effective distance versus ringdown frequency $f_0$ for simulated signals found in coincidence. Simulated signals found in triple coincidence are marked as green x's, simulated signals found in double coincidence are shown as red dots and those signals found in double coincidence because the third detector was vetoed are circled in black.}
\label{fig:mf_allvH}
\end{center}
\end{figure}

The signal is modeled by adding simulated signals in software to the data stream and running the pipeline in the same manner as one would with the real data. Fig.~\ref{fig:mf_allvH} displays a plot of Hanford effective distance (the distance to an optimally located and oriented source) as a function of frequency for simulated signals that were found in all three detectors and in all combinations of two detectors (H1H2, H1L1, H2L1). The plot shows that the search is most sensitive to ringdown signals occurring in the \unit{70}{Hz}--\unit{140}{Hz} band, where detector noise is lowest. 

The background, or false alarm rate, is estimated by shifting two data streams in time with respect to one-another and running the pipeline described in section~\ref{sec:pipeline}. The L1 data stream is slid fifty times in multiples of \unit{$\pm10$}{s} and H2 is slid fifty times in multiples of \unit{$\pm5$}{s}. As the time steps are much larger than the longest possible delay between detectors in receiving a real gravitational wave signal, any coincidences found cannot be due to gravitational waves, and are therefore considered false coincidences. We use these as estimates of the background due to false in-time coincidence. 

As a final sanity check before the search is unblinded we look at approximately one tenth of the data to ensure that the false alarm rate is consistent with our background studies. In order to avoid any potential bias in the tuning procedure affecting the upper limits, these data are excluded from the upper limit calculation.


\section{Results}

\begin{figure}[ht]
\centering
\begin{center}
\includegraphics[scale=0.45]{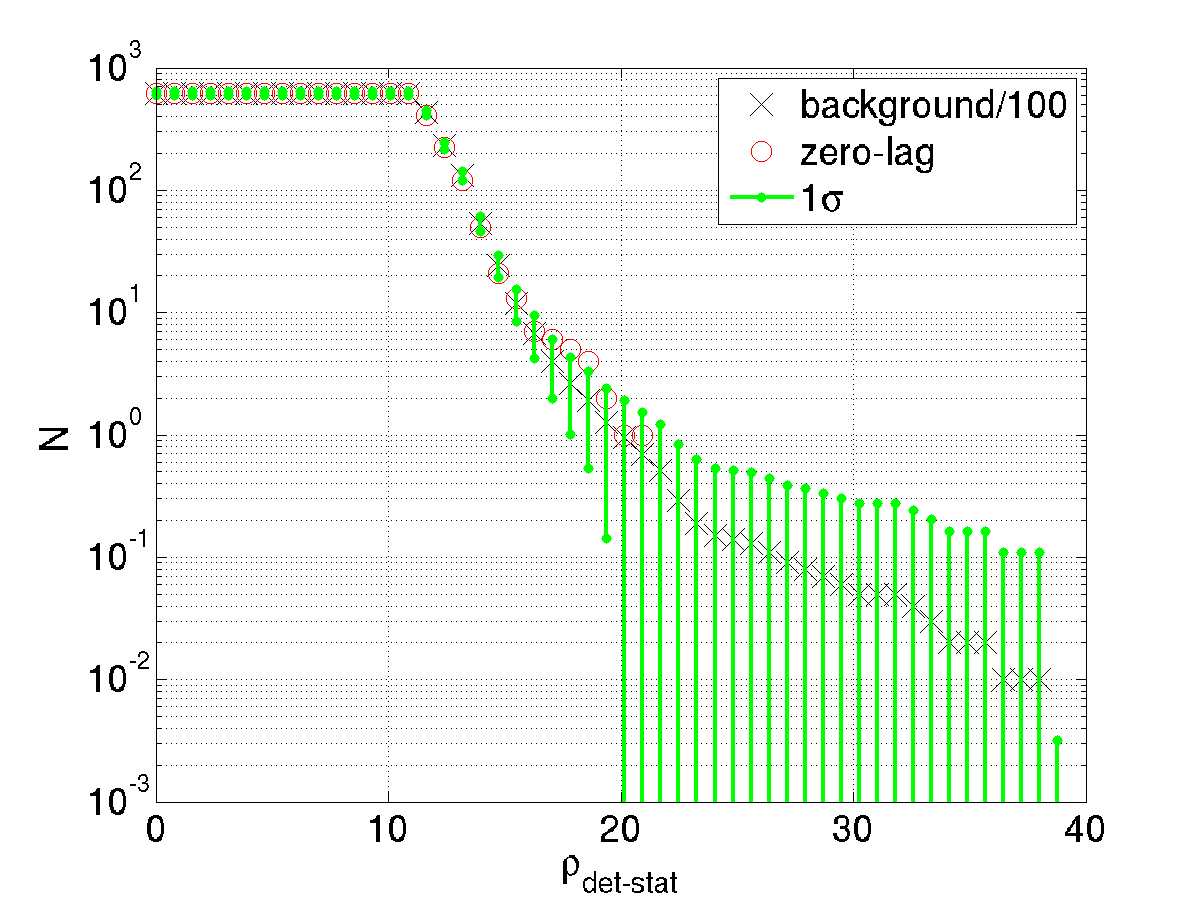}
\caption{Cumulative histogram of coincident triggers (red circles) and an estimate of the background (black x's): H1L1 doubles in triple time.}
\label{fig:bgintdinttH1L1}
\end{center}
\end{figure}

Once the tuning is complete and all thresholds and cuts are finalized the pipeline described above is run on the full data set. Unblinding the search with the cuts and thresholds described above revealed no triple coincident events. A number of double coincidences were found to satisfy the constraints; however, these events are consistent with background as shown in Fig.~\ref{fig:bgintdinttH1L1}. The loudest candidate events were subjected to further investigation, and in each case there was insufficient evidence that the coincidence could not be attributed to noise in the detector.

\begin{figure}[hbtp]
\includegraphics[width=0.5\textwidth]{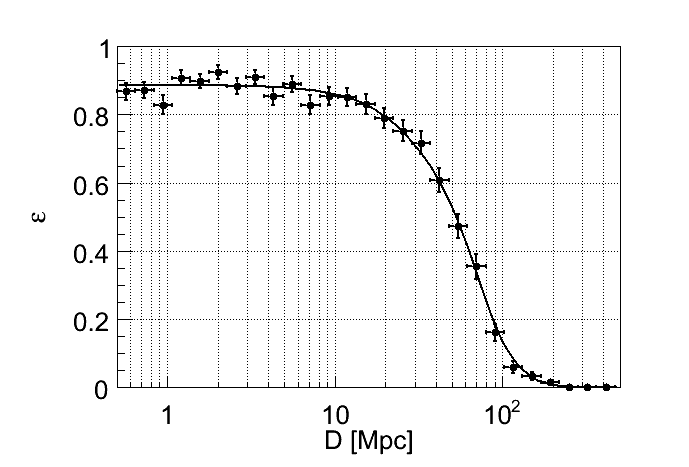}
\caption{The efficiency of detecting simulated signal injections in triple coincidence in the \unit{70}{Hz}--\unit{140}{Hz} band, fit to a sigmoid.}
\label{fig:eff}
\end{figure}

We calculate an upper limit on the rate of ringdowns for a given population of black holes using simulated signals to evaluate the efficiency of the search, $\varepsilon(r)$, defined as the fraction of simulated signals detected in the analysis, as a function of physical distance. The simulated signals were uniform in orientation and direction. Given the expected high false alarm rate in double coincidence from background studies, we set an upper limit from times when all three detectors were recording data, a total of \unit{$T=0.0375$}{years}. Fig.~\ref{fig:mf_allvH} shows that the efficiency is strongly dependent on the ringdown frequency $f_0$, and thus for the purpose of setting an upper limit we restrict the calculation to the most sensitive frequency band, \unit{70}{Hz}--\unit{140}{Hz}, corresponding to black hole masses in the range \unit{85}{$M_\odot$}--\unit{390}{$M_\odot$}. We added 5701 simulated signals in the frequency range of \unit{70}{Hz}--\unit{140}{Hz} and between \unit{0.5}{Mpc} and \unit{$10^3$}{Mpc} in distance to the data, over ten runs. Of these, 3010 were recovered in triple coincidence. Fig.~\ref{fig:eff} displays the efficiency as a function of distance for the frequency band of interest. (Note that the efficiency is not equal to 1 at nearby distances because we apply vetoes and treat these as a loss of efficiency rather than a loss in live-time.)  
From the efficiency we can calculate the effective volume we are sensitive to, $V_{\textrm{eff}}$
\begin{equation}
V_{\textrm{eff}}=4 \pi \int \varepsilon(r) r^2 \textrm{d}r.
\end{equation}
For this band \unit{$V_{\textrm{eff}}=2.6 \times 10^6$}{Mpc$^3$}. This corresponds to a typical distance to a source of $\approx 85$ Mpc. The 90\% confidence upper limit on the rate is given by~\cite{loudestGWDAW03}
\begin{equation}
R_{90\%}=\frac{2.303}{T V_{\textrm{eff}}},
\end{equation}
which, for the \unit{70}{Hz}--\unit{140}{Hz} band is $\mathunit{2.4\times10^{-5}}{\textrm{yr}^{-1}\textrm{Mpc}^{-3}}$. 
Even though our knowledge of the population of intermediate mass black holes is poor, as discussed in section \ref{sec:intro}, we do know that the formation of stars in general scales with the blue-light luminosity emitted by galaxies, and as it is expected that the rate of binary coalescence (including ringdown) follows the rate of star formation, we can work under the assumption that the rate of ringdowns also scales with blue light luminosity~\cite{Phinney:1991ei}. We introduce the cumulative blue luminosity, $C_L$, observable within the range of the search, 
\begin{equation}
C_L=\rho_LV_{\textrm{eff}}.
\end{equation}
$C_L$ has units of $L_{10}$, where  $L_{10}=10^{10}L_{B,\odot}$ and $L_{B,\odot}$ is the B-band solar luminosity (1 MWEG is equivalent to \unit{1.7 $L_{10}$}), and \unit{$\rho_L=(1.98 \pm 0.16) \times 10^{-2} L_{10}$}{Mpc$^{-3}$} is the average blue luminosity density~\cite{LIGOS3S4Galaxies}. The 90\% confidence upper limit on the rate in these units is given by
\begin{equation}
R_{90\%}=\frac{2.303}{T C_L},
\end{equation}
which evaluates to $\mathunit{1.2\times10^{-3}}{\textrm{yr}^{-1}L_{10}^{-1}}$.

Due to our lack of knowledge about the population of black holes we assign no error to the astrophysical source population. Similarly, we assign no error to the waveform: comparison with numerical relativity results has shown that exponentially-damped-sinusoid templates perform well at detecting the signal and characterizing the black hole parameters~\cite{2009CQGra..26p3001B}. We limit ourselves to evaluating systematic errors associated with the experimental apparatus and analysis method. The only error that we associate with the former is calibration of the data, and with the latter is with the limited number of Monte Carlo simulations used to evaluate the efficiency.
Errors in the calibration can cause the SNR of a signal to be incorrectly quantified, thereby introducing inaccuracies in the distance. As the efficiency is a function of distance and frequency, care has to be taken to adjust the efficiency curve appropriately. The fractional uncertainty in amplitude was found from calibration studies \cite{Dietz:2006} to be 5\%. This results in an error of \unit{$3.5\times10^5$}{Mpc$^3$} in $V_{\textrm{eff}}$. 
The second source of error is due to the limited number of simulated signals in our Monte Carlo (MC) simulations to evaluate the efficiency. Assuming binomial errors we calculate the variance of the efficiency $\sigma^2_{\textrm{MC}}$,
 which corresponds to an error in $V_{\textrm{eff}}$ of $2.3\times10^5$ Mpc$^3$. These errors are summed in quadrature, and multiplied by 1.64 to give a 90\%  confidence interval of \unit{$6.9\times10^5$}{Mpc$^3$}.  To obtain an upper limit we apply a downward excursion to the effective volume and obtain \unit{$R_{90\%} = \unit{$3.2\times 10^{-5}$}{yr$^{-1}$ Mpc$^{-3}$} = 1.6 \times 10^{-3}$}{yr$^{-1} L_{10}^{-1}$}.


\section{Conclusion}
We have performed the first search for gravitational waves signals from black hole ringdowns in LIGO data, and demonstrated with simulated signals that this pipeline is an effective method of detecting such signals in coincidence between all three LIGO detectors. The search encompassed black holes in the mass range of \unit{10}{$M_\odot$}--\unit{500}{$M_\odot$}, the regime of intermediate mass black holes, with spins ranging from 0 to 0.994. No gravitational-wave events were found, and an upper limit of \unit{$R_{90\%}=1.6 \times 10^{-3}$}{yr$^{-1} L_{10}^{-1}$} was placed on the rate of ringdowns from black holes formed from binary mergers, in the mass range of \unit{85}{$M_\odot$}--\unit{390}{$M_\odot$}. 

A search for black hole ringdowns in data from the fifth LIGO science run is currently underway. This two year long science run was the first at LIGO design sensitivity. With the increase in sensitivity of the LIGO detectors between the two runs, the ringdown horizon distance of S5 is greater than that of S4, allowing access to a greater number of potential sources from which a detection could be made. A further increase in sensitivity will come with Advanced LIGO, allowing us to detect compact binary coalescence to cosmological distances, and the improved sensitivity at lower frequency will make us sensitive to black holes with masses up to \unit{2000}{$M_\odot$} or higher.


\acknowledgments

The authors gratefully acknowledge the support of the United States
National Science Foundation for the construction and operation of the
LIGO Laboratory and the Science and Technology Facilities Council of the
United Kingdom, the Max-Planck-Society, and the State of
Niedersachsen/Germany for support of the construction and operation of
the GEO600 detector. The authors also gratefully acknowledge the support
of the research by these agencies and by the Australian Research Council,
the Council of Scientific and Industrial Research of India, the Istituto
Nazionale di Fisica Nucleare of Italy, the Spanish Ministerio de
Educaci\'on y Ciencia, the Conselleria d'Economia, Hisenda i Innovaci\'o of
the Govern de les Illes Balears, the Scottish Funding Council, the
Scottish Universities Physics Alliance, The National Aeronautics and
Space Administration, the Carnegie Trust, the Leverhulme Trust, the David
and Lucile Packard Foundation, the Research Corporation, and the Alfred
P. Sloan Foundation.
This paper was assigned LIGO document number LIGO-P080093.


\bibliography{iulpapers}

\end{document}